\documentclass[preprint]{revtex4-1}
\usepackage{graphicx,color}  
\usepackage{subfigure} 
\usepackage{bm}        
\usepackage{amssymb}   
\usepackage{amsmath}   
\usepackage[nice]{nicefrac}
\usepackage{exscale}
\usepackage{xcolor}

\definecolor{darkred}{rgb}{0.90,0.2,0.2}

\usepackage[plainpages=false,pdfpagelabels,colorlinks=true,linkcolor=red,urlcolor=blue,citecolor=blue,pdftitle={},pdfauthor={},pdfdisplaydoctitle=true,pdfduplex=DuplexFlipLongEdge]{hyperref}

\begin{document}

\title{Observation of Dynamical Fermionization}

\author{Joshua M. Wilson}
\affiliation{Department of Physics, The Pennsylvania State University, University Park, PA 16802, USA}

\author{Neel Malvania}
\affiliation{Department of Physics, The Pennsylvania State University, University Park, PA 16802, USA}

\author{Yuan Le}
\affiliation{Department of Physics, The Pennsylvania State University, University Park, PA 16802, USA}

\author{Yicheng Zhang}
\affiliation{Department of Physics, The Pennsylvania State University, University Park, PA 16802, USA}

\author{Marcos Rigol}
\affiliation{Department of Physics, The Pennsylvania State University, University Park, PA 16802, USA}

\author{David S. Weiss}
\affiliation{Department of Physics, The Pennsylvania State University, University Park, PA 16802, USA}

\begin{abstract}
We observe dynamical fermionization, where the momentum distribution of a Tonks-Girardeau (T-G) gas of strongly interacting bosons in 1D evolves from bosonic to fermionic after its axial confinement is removed.  The asymptotic momentum distribution after expansion in 1D is the distribution of rapidities, which are the conserved quantities associated with many-body integrable systems. Rapidities have not previously been measured in any interacting many-body quantum system. Our measurements agree well with T-G gas theory. We also study momentum evolution after the trap depth is suddenly changed to a new non-zero value. We observe the predicted bosonic-fermionic oscillations and see deviations from the theory outside of the T-G gas limit.
\end{abstract}

\maketitle

Integrable many-body quantum systems have been extensively studied theoretically since Bethe solved the 1D Heisenberg model in 1931~\cite{Bethe1931}. The theoretical appeal of these systems stems from the deep symmetries they exhibit and the fact that it is possible to exactly solve for their many-body wavefunctions~\cite{sutherland_book_04}. In a development that would have surprised the mathematical physicists who started the field, over the last 20 years there have been more than a dozen experimental implementations of very nearly integrable models. Systems of bosons~\cite{cazalilla_citro_review_11}, spins~\cite{cazalilla_citro_review_11} and fermions~\cite{guan_batchelor_review_13} have been realized, using a range of ultracold atom, trapped ion and condensed matter techniques. All these integrable many-body systems are characterized by infinite sets of conserved quantities, known as the distributions of rapidities. The rapidities embody what makes integrable systems special, including the fact that they do not reliably thermalize under unitary dynamics (see Ref.~\cite{calabrese_essler_review_16} for a recent set of reviews on this topic). The many-body character of rapidities in interacting integrable systems makes their distributions difficult, if not impossible, to extract from equilibrium measurements.  However, when the particles in an integrable system are allowed to expand in one dimension, the interparticle interactions vanish asymptotically and the momentum distribution approaches the distribution of rapidities~\cite{sutherland_98, rigol_muramatsu_05a, minguzzi_gangardt_05, bolech_meisner_12, campbell_gangardt_15, mei_vidmar_16}.

We have performed just such an expansion measurement with a Lieb-Liniger gas~\cite{Lieb_Liniger_63}, an integrable system of 1D bosons with contact interactions. We operate in the Tonks-Girardeau (T-G) gas limit~\cite{kinoshita_wenger_04, paredes_widera_04, kinoshita_wenger_05}, where the interactions are very strong~\cite{cazalilla_citro_review_11}. The many-body wavefunction of the T-G gas is the same as that of a non-interacting Fermi (NIF) gas, to within an absolute value. All local properties are the same for the two gases, but non-local properties, like the momentum distributions, are different. Remarkably, the distribution of rapidities of the T-G gas is the momentum distribution function of the NIF~\cite{cazalilla_citro_review_11}. By observing the T-G gases' bosonic momentum distribution function dynamically ``fermionize''~\cite{rigol_muramatsu_05a}, we have directly measured the distribution of rapidities in this many-body interacting quantum system, thus bringing these theoretical constructs into the realm of experiment. Our experimental results for time-of-flight (TOF) measurements are in almost perfect agreement with exact numerical calculations. We have also measured other momentum distribution dynamics after quenches to different non-vanishing trap strengths~\cite{minguzzi_gangardt_05}.

\begin{figure}[!h]
	\includegraphics[width=12cm]{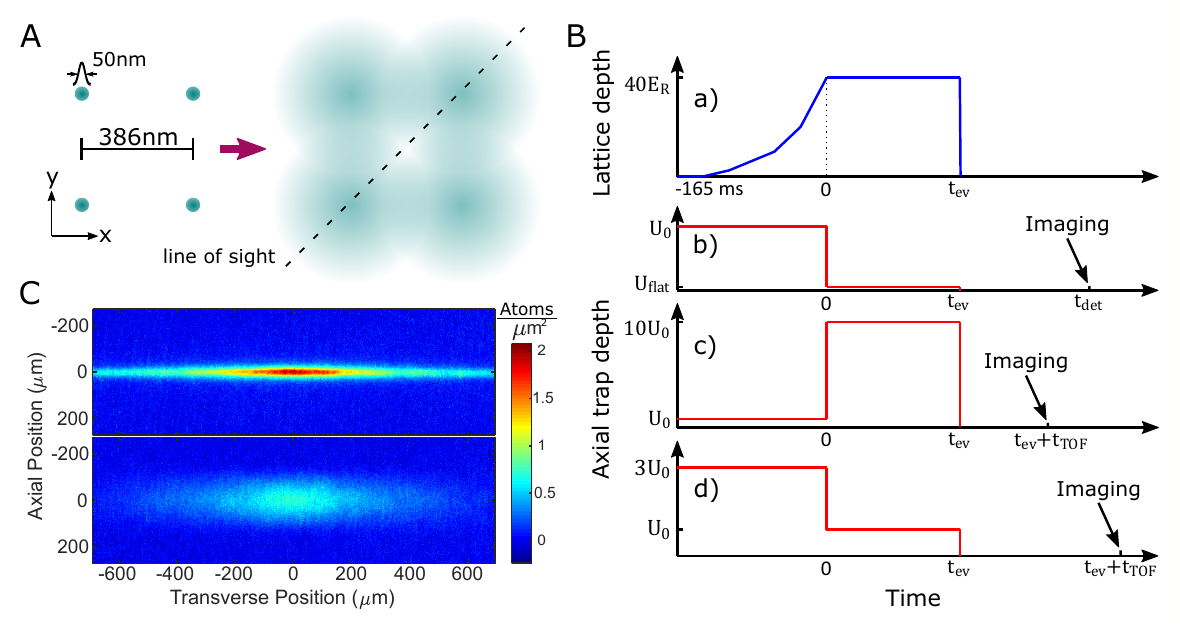}
	\caption{Timing and measurement. ({\bf A}) Schematic of the momentum measurement. Atoms are initially confined in a 2D optical lattice of 1D tubes. When the 2D lattice is shut off, rapid transverse expansion reduces the density, taking away interaction energy and allowing a good momentum measurement. Absorption imaging is done along the line of sight. ({\bf B}) Timing diagram, not to scale horizontally. (a) The lattice depth as a function of time. (b) The axial trap depth as a function of time for the dynamical fermionization measurement. At $t=0$ the depth is suddenly lowered to cancel out the residual anti-trap due to the lattice beams. All traps are shut off at a variable $t_{ev}$, and imaging occurs at a fixed $t_{det}$ (relative to $t=0$). (c) and (d) The axial trap depth as a function of time for the Bose-Fermi oscillation experiments. The axial trap depth is suddenly changed at $t=0$ and the atoms evolve in the new trap for a variable $t_{ev}$. The absorption image is taken at $t=t_{TOF}+t_{ev}$. ({\bf C}) Absorption images for $t_{ev}=0$ (upper image) and $t_{ev}=15$~ms (lower image), after quenching to a flat potential. The images are averages over 30 shots. Sudden lattice shut-off makes the atoms expand rapidly transversely. The 1D TOF distributions (in the $z$-direction, vertical in the images) are obtained by integrating the images transversely.}\label{fig:sketch}
\end{figure}

The momentum distributions of equilibrium 1D Bose gases have been previously measured with TOF, Bragg spectroscopy, observation of phase fluctuations, and with momentum focusing techniques ~\cite{cazalilla_citro_review_11, bloch_dalibard_review_08}. These measurements have all been initiated by shutting off both axial and transverse trapping, which precludes the expansion in 1D that is required for a rapidity measurement. In our experiment we can remove the axial potential without affecting the transverse trapping that makes the system 1D, thus allowing for free expansion in 1D. We initiate the momentum measurement at controllable times, $t_{ev}$, during the 1D expansion by suddenly shutting off the transverse trapping (see Fig.~\ref{fig:sketch}A). The wavefunctions rapidly expand transversely, which dramatically decreases their interaction energy before the axial wavefunction appreciably changes. After a long TOF, the spatial distribution approaches the momentum distribution at $t_{ev}$ (see Fig.~\ref{fig:sketch}B).

The experiment starts with a BEC of $10^5$ $^{87}$Rb atoms in the $F=1$, $m_F=1$ state trapped in a crossed dipole trap, around which we slowly turn on a blue-detuned 2D optical lattice to a depth of $40E_R$, where $E_R=\nicefrac{(\hbar k)^2}{2m}$ is the recoil energy, $m$ is the Rb mass, and $k=\nicefrac{2\pi}{772\,{\text nm}}$ is the lattice wavevector~\cite[see Supplemental Information (SI)]{kinoshita_wenger_04, kinoshita_wenger_05}. The atoms end up trapped in a 2D array of nearly identical ``tubes'' with negligible tunneling among them. The number of particles per tube varies from 26 to 0 (see SI). The axial trapping frequency is approximately the same in all the occupied tubes, $\nicefrac{\omega_z}{2\pi}=18.1\pm0.36~Hz$. The Lieb-Liniger model that describes these 1D gases is characterized by the dimensionless coupling strength $\gamma$~\cite{cazalilla_citro_review_11}. For large $\gamma$ there are strong correlations among the single-particle wavefunctions, since it is too energetically costly for them to significantly overlap. In our tubes, $\gamma=\nicefrac{4.44}{n_{1D}}$, where $n_{1D}$ is the local 1D density in $\mu$m$^{-1}$~\cite{olshanii_98}. With our initial trapping parameters the weighted average $\gamma$ is 8.5, and the smallest $\gamma$ is 4.2 at the center of the central tube. Our theoretical analysis assumes the T-G gas limit of $\gamma \rightarrow \infty$.

\begin{figure}[!tp]
	\includegraphics[width=12cm]{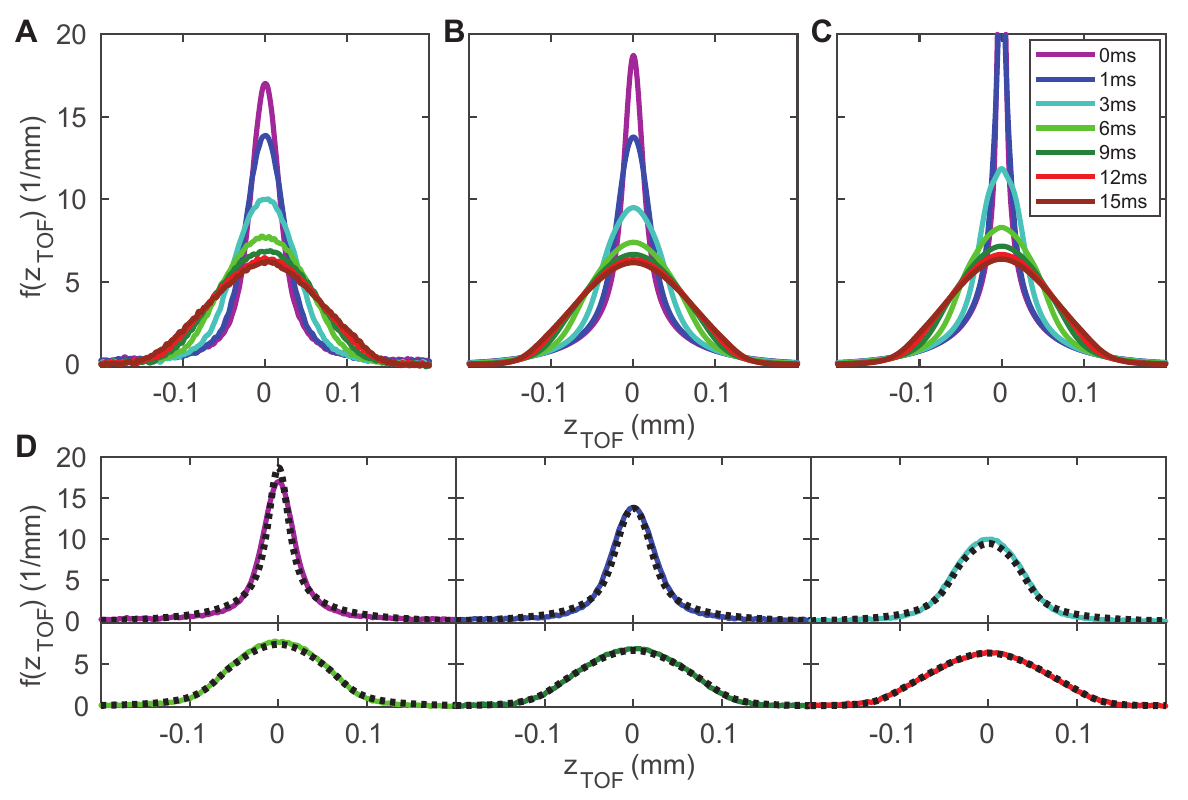}
	\caption{Dynamical fermionization. ({\bf A}) Normalized experimental axial TOF distributions for a range of $t_{ev}$. Each profile is an average of 30 implementations. By 15~ms the shape has asymptoted. ({\bf B}) Numerical simulation of the experiment in the T-G limit, with no free parameters. ({\bf C}) The corresponding numerical simulation of the momentum distribution functions (rescaled by the $t_{DET}$). ({\bf D}) Experimental distributions for the first six times shown in A (colored curves), separately compared to the corresponding theoretical curves from B (dotted black lines). After 12~ms the theory and experiment are essentially indistinguishable, and very close to the theoretical momentum distributions. }\label{fig:ferm}
\end{figure}

To demonstrate dynamical fermionization, we suddenly reduce the depth of the crossed dipole trap at $t=0$ so that, when combined with the weak axial anti-trap due to the blue detuned 2D lattice, there is an approximately flat potential over an axial range of about $40\,\mu$m (see Fig.~\ref{fig:AntiTrap}A). After a variable $t_{ev}$ we turn off the 2D lattice in $32~\mu$s, which yields a TOF distribution that is barely distinguishable from what is obtained after a sudden shutoff (see Fig.~\ref{fig:Shutoff}B). The slower turn off reduces the transverse expansion (see Fig~\ref{fig:Shutoff}A), which allows for a longer TOF before the atoms spread to a region where gravity is not well-cancelled by our magnetic field gradient .  At $t_{det} =70$~ms we take absorption images of the atoms (see Fig.~\ref{fig:sketch}C) and integrate over the transverse direction to obtain the TOF 1D distributions. The results are shown in Fig.~\ref{fig:ferm}A. The initially peaked ``bosonic'' TOF distribution smoothly deforms and approaches a rounded ``fermionic'' TOF distribution over the first 12~ms (see also Fig.~\ref{fig:AntiTrap}B which shows the full width at half maximum (FWHM) of the TOF distributions). In the 12 ms over which the distribution has mostly fermionized, the axial spatial extent of the atoms (before TOF) grows from a FWHM of 22 to 42 $\mu$m. When $t_{ev}>15$~ms the atoms have expanded to where the axial potential is insufficiently flat and just starts to effect the TOF distribution.

We perform numerical simulations of our experiment using the continuum limit of a lattice hard-core boson model~\cite{xu_rigol_15}, which incorporates all the experimental details, including the initial size, the evolution up to $t_{ev}$, the TOF, the instrumental resolution ($4.8~\mu$m), and the sum over tubes (see SI). The results are shown in Fig.~\ref{fig:ferm}B. Figure~\ref{fig:ferm}D shows direct comparisons between individual experimental curves (solid lines) and their numerical counterparts (black dotted lines). With no free parameters, the simulations match the experimental results well, particularly in the asymptotic limit. The agreement at long times suggests that the T-G gas model is sufficient for our finite $\gamma$ system. The small discrepancies at earlier times are probably due to the non-zero initial temperatures in the experiments, which are known to strongly affect the height of the zero momentum peak in the T-G limit~\cite{rigol_05}.

To see how the initial size, instrumental resolution and finite $t_{det}$ affect these results, we show the evolution of the theoretical momentum distributions in Fig.~\ref{fig:ferm}C. At small $t_{ev}$ these factors broaden the measured widths, but as the asymptotic limit is approached the TOF distributions are nearly identical to the actual momentum distributions. We have thus measured the distribution of rapidities, the first time these quantities have been observed in a many-body quantum system.

We have also studied the dynamics after suddenly changing the depth of the axial trap. Related quenches have previously been studied in the weakly interacting regime~\cite{PhysRevLett.113.035301}. A numerical simulation in the T-G limit has shown that, for harmonic traps, a sudden 10-fold reduction in axial trapping frequency leads to the surprising behavior that the momentum distribution oscillates between bosonic and fermionic shapes~\cite{minguzzi_gangardt_05}. The initial change to a fermionic shape is easily understood as approximately dynamical fermionization. What is more remarkable and counterintuitive is the return at $T/2$ to a bosonic distribution with a height and width changed by a factor of the ratio of the oscillation frequencies, $r=\omega_f/\omega_0$, $f(p,T/2)=r f(r p, 0)$. In the second half of the period the distribution evolves through the fermionized distribution back to the original bosonic one.

\begin{figure}[!tp]
	\includegraphics[width=12cm]{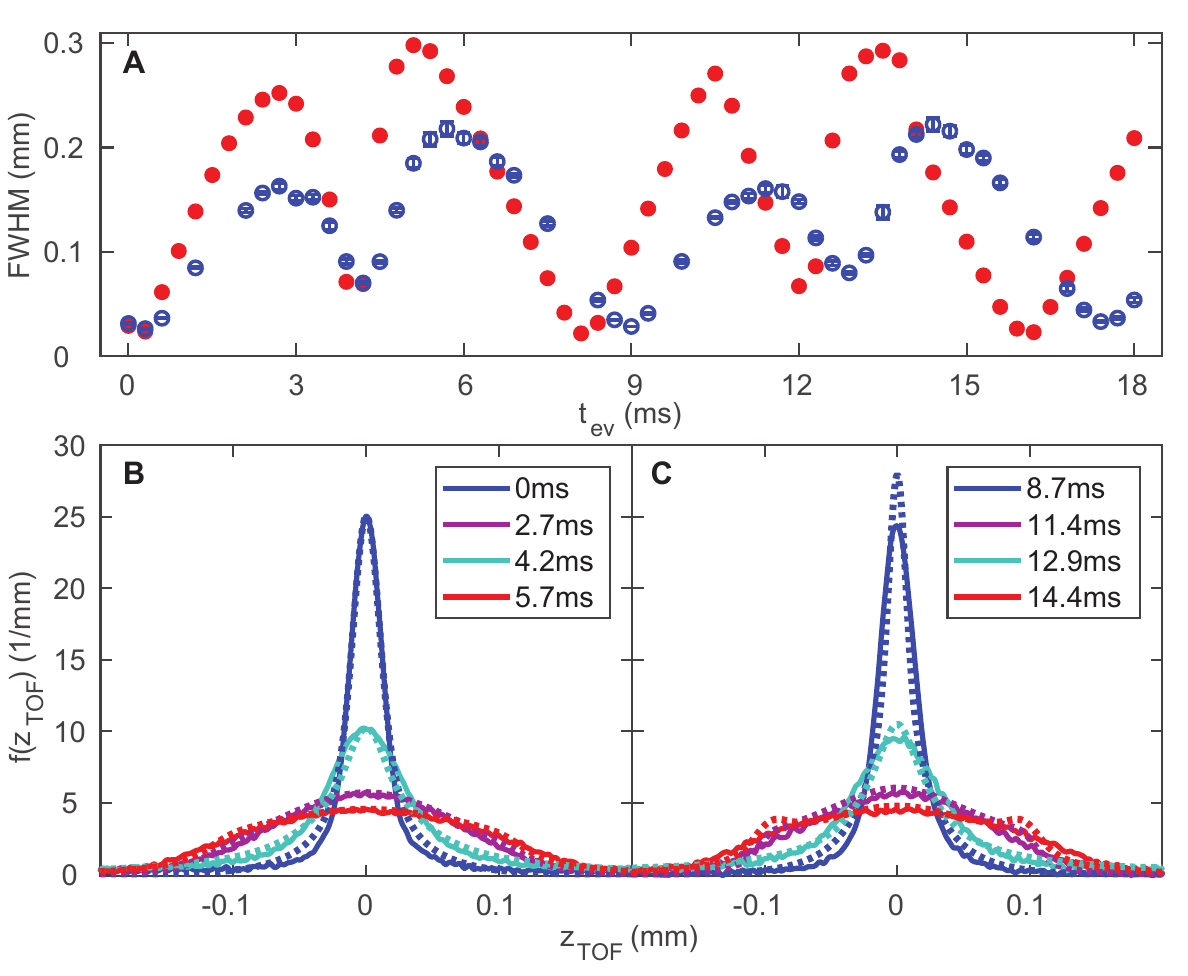}
	\caption{Bose-Fermi oscillations (quench from low to high $\omega_z$). ({\bf A}) FWHM as a function of time after the quench to a 10 times deeper axial trap. The blue points are from the experiment, with standard error bars from an average of 5 to 14 shots (see SI). The red points are from the T-G gas theory. For a few points in the second period the center of the distribution is not the maximum (see Fig~\ref{fig:EBP}); in those cases we still define the half maximum relative to the center point. We attribute the difference in oscillation period to finite $\gamma$ in the experiment. ({\bf B}) TOF distributions associated with the extrema of the first oscillation cycle. The experimental curves are solid, and the corresponding theoretical curves are dotted. The shapes at the minima (blue and teal) are bosonic, with small differences associated with finite initial sizes. The shapes at the maxima (purple and red) are fermionic, like the asymptotic dynamical fermionization distribution. The theoretical curves have been rescaled to better compare the shapes to the experimental curves. ({\bf C}) TOF distributions associated with the extrema of the second oscillation cycle. The shapes at the minima (blue and teal) are bosonic. The experimental curves at the maxima (purple and red) are fermionic, but the theoretical curves have small sidelobes that are associated with the axial trap anharmonicity. We suspect their absence in the experiment is a consequence of the smaller $\gamma$ (see main text).}\label{fig:osc}
\end{figure}

We experimentaly perform quenches both to a 10 times deeper trap and to a 3 times shallower trap. The former makes initial size effects less important, so that the TOF distributions better approximate the momentum distributions. However, quenching to a deeper trap decreases $\gamma$ to an average of near 2 at $T/2$ (see SI), which worsens the T-G gas approximation.  Higher densities require that we shut off the lattice as fast as possible to prevent axial evolution while the interaction energy is being removed, which in turn limits the available TOF time, $t_{TOF}$, to 40~ms (see SI). We first characterize the TOF distributions of the evolving gas in a shape-agnostic way by plotting the FWHM versus time over the first two periods, as shown by the blue points in Fig.~\ref{fig:osc}A. The corresponding T-G gas theory curves are shown by the red points. The theoretical period is $\sim 9\%$ shorter than in the experiment. The longer experimental period is expected, based on the known functional dependence of the ratio of breathing to dipole oscillation frequencies, which varies from 2 to $\sqrt{3}$ when $\gamma$ goes from $\infty$ to 0 \cite[see SI]{menotti_stringari_02, moritz_stoferle_03}.

The solid lines in Figs.~\ref{fig:osc}B and \ref{fig:osc}C show the experimental TOF distributions near the peaks and valleys from Fig.~\ref{fig:osc}A (see Fig.~\ref{fig:EBP} for the shapes at other times). The dotted lines in Figs.~\ref{fig:osc}B and \ref{fig:osc}C are from corresponding theory curves, with the heights and widths rescaled for easier comparison of the shapes (see SI). Focusing on the first period, the salient point is that the theory and experimental shapes evolve in the same way. They are bosonic at 0 and $T/2$, with experiemental and theoretical widths that are within $6\%$ of each other~\cite{minguzzi_gangardt_05}. They are fermionic at the FWHM peaks and at the surrounding points (see Fig.~\ref{fig:EBP}). The asymmetry about $T/2$ is a finite size effect. The fact that the fermionic FWHMs are smaller in the experiment than the theory is a consequence of finite $\gamma$ in the former.

The experimental shapes are almost identical between the first and second periods, highlighting the lack of damping in this integrable evolution. The theoretical shapes, however, are slightly different near the FWHM peaks in the second period, showing flattening at the top and side peaks. We use the width and amplitude rescaling from the first cycle on the second cycle theory curves. The new features in the theory curves result from the Gaussian trap's small deviation from harmonicity; they are absent when we use a harmonic trap for the calculation (see Fig.~\ref{fig:TheoryComparisons}). We suspect that the absence of these features in the experiment results from the reduced $\gamma$. A similar discrepancy between experiments and $\gamma\rightarrow\infty$ theory was seen in Ref. ~\cite{Vidmar_15}.

\begin{figure}[!tp]
	\includegraphics[width=12cm]{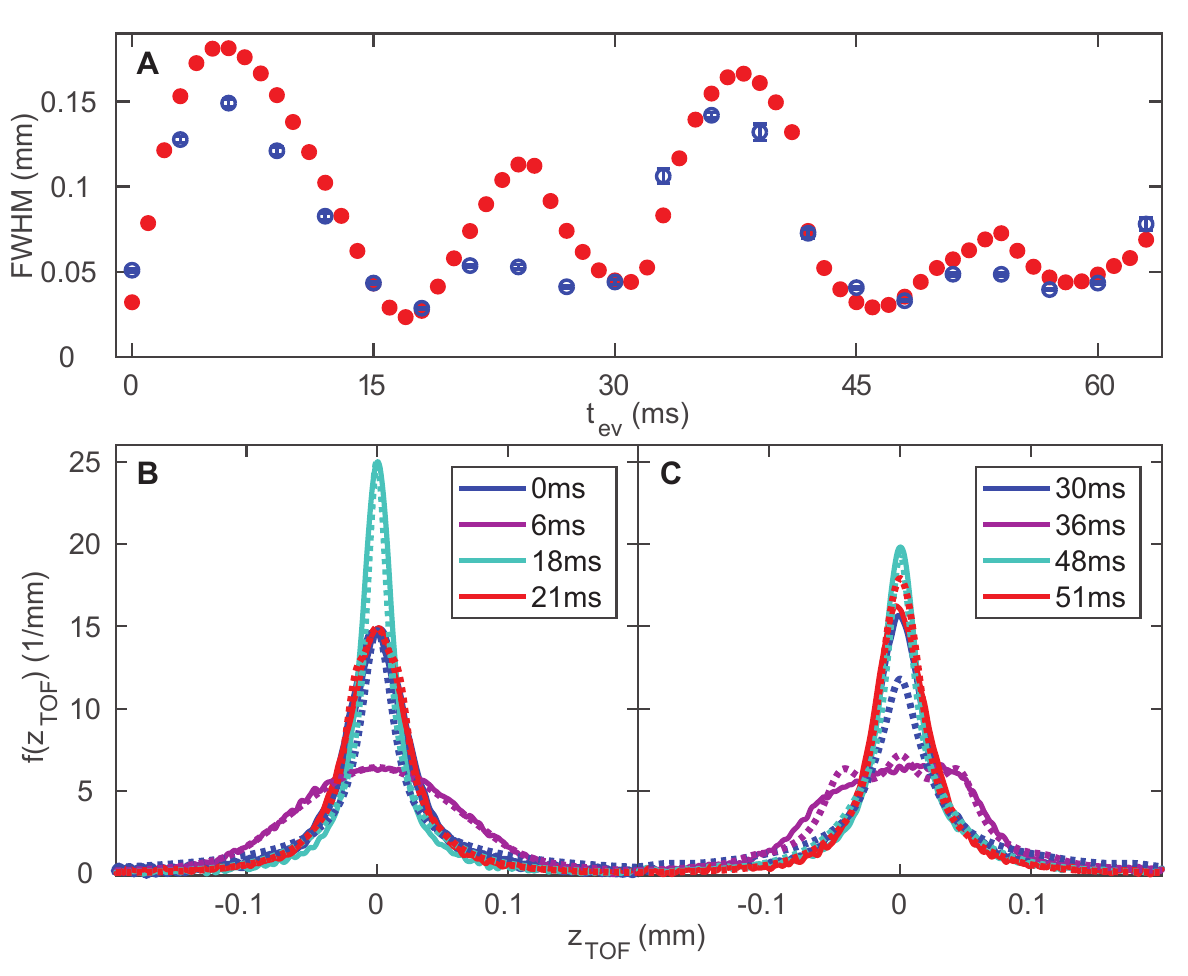}
	\caption{Bose-Fermi oscillations (quench from high to low $\omega_z$). ({\bf A}) FWHM as a function of time after the quench to a 3 times shallower axial trap. The blue points are from the experiment, with standard error bars from an average of 10 shots. The red points are from the T-G gas theory. ({\bf B}) TOF distributions associated with the extrema of the first oscillation cycle. The experimental curves are solid, and the corresponding theoretical curves are dotted. The shapes at the minima (blue and teal) are bosonic, with small differences associated with finite initial sizes. The shapes at the maxima (purple and red) are fermionic, like the asymptotic dynamical fermionization distribution. The theoretical curves have been rescaled to better compare the shapes to the experimental curves. ({\bf C}) TOF distributions associated with the extrema of the second oscillation cycle. The shapes at the minima (blue, teal and black) are bosonic. The experimental curves at the maxima (purple and red) are fermionic. Both theoretical and experimental curves have distorted shapes associated with the axial trap anharmonicity.}\label{fig:osc2}
\end{figure}

In the quench to a shallower trap $\gamma$ increases from 4.4 to $\sim$6.7 during the oscillation. The observed period matches that of the T-G gas theory (see Fig.~\ref{fig:osc2}A), possibly because two small frequency shifts cancel (see SI). The first cycle shapes are similar to those in the other quench (see Figs.~\ref{fig:osc2}B and ~\ref{fig:EBP2}). In the second cycle, we observe a flattening in the experimental distribution near the FWHM peaks in both the experiment and theory (see Figs.~\ref{fig:osc2}C and ~\ref{fig:EBP2}). $\gamma$ is apparently large enough that this effect of anharmonicity is not completely smoothed out, but still far enough from $\infty$ to suppress the FWHM peaks.

The technique presented here can also be used to measure rapidity distributions, and to explore expansion dynamics of density and momentum distributions, in intermediate-$\gamma$ 1D Bose gases. This is complementary to what is accessible in atom-chip experiments~\cite{schemmer_bouchoule_19}, and provides a broad testing ground for the recently developed generalized hydrodynamics theory~\cite{bertini_collura_16, castro_doyon_16}. Our technique can also be applied to measuring rapidity distributions and momentum dynamics after more complex quenches, like those in quantum Newton's cradles~\cite{kinoshita_wenger_06, tang_kao_18}, recently studied theoretically using generalized hydrodynamics~\cite{caux_doyon_19}. It can be applied to 1D lattice models, such as the 1D Fermi-Hubbard model~\cite{bolech_meisner_12, mei_vidmar_16}. Knowledge of the rapidity distributions, together with the theoretical tools that have been developed in the field of integrable quantum systems, allows predictions of all aspects of integrable quantum systems, including correlation functions and dynamics.

\newpage
\section*{FIGURES: 1-4}
\newpage
\newpage
\newpage
\newpage

\clearpage
\bibliographystyle{apsrev4-1}
\bibliography{references}
\section*{Acknowledgments}
	This work was supported by the National Science Foundation under Grant No.~PHY-1707482 (Y.Z. and M.R.) and No.~PHY-1707576 (D.S.W., J.M.W., N.M., and Y.L.) and by the U.S. Army Research Office No.~W911NF-16-0031-P00005 (D.S.W., J.M.W., N.M., and Y.L.). The computations were carried out at the Institute for CyberScience at Penn State.

\clearpage
\newpage

\setcounter{figure}{0}
\setcounter{equation}{0}
\setcounter{page}{1}

\renewcommand{\thetable}{S\arabic{table}}
\renewcommand{\thefigure}{S\arabic{figure}}
\renewcommand{\theequation}{S\arabic{equation}}

\renewcommand{\thesection}{S\arabic{section}}

\onecolumngrid

\begin{center}
	
	{\large \bf Supplemental Material:\\ Observation of Dynamical Fermionization}
	
	\vspace{0.3cm}
	
	J. M. Wilson, N. Malvania, Y. Le, Y. Zhang, M. Rigol, D. S. Weiss
	
	{\it Department of Physics, The Pennsylvania State University, University Park, PA 16802, USA}
	
\end{center}

\onecolumngrid

\label{pagesupp}

\section*{Supplemental Information}

\subsection{Experimental design}
We create a nearly pure BEC by evaporatively cooling ${}^{87}Rb$ atoms in the $F=1,\,m_f=1$ ground state \cite{kinoshita_05_BEC}. Two beams of red-detuned $1064$ nm light, each with $2.2$ mW of power, form a crossed dipole trap with a $59$ $\mu$m waist and provide confinement with an axial trapping frequency of $19$ Hz. A pair of Helmholtz coils generates a $30$ G/cm magnetic field gradient to levitate the spin-polarized atoms against gravity. We adiabatically ramp up a 2D optical lattice around the BEC to trap the atoms in arrays of tubes \cite{kinoshita_wenger_04}. The lattice is formed by two orthogonal beams of blue-detuned $772$ nm light with $420$ $\mu$m waists.

We ramp up the lattice in four linear segments: from $0-2.5\,E_R$ in $35$ ms, $2.5-10\,E_R$ in $60$ ms, $10-20\,E_R$ in $35$ ms, and finally $20-40\,E_R$ in $35$ ms. The sequence of ramps keeps the amplitude of any non-adiabatic breathing excitation to within $5\%$ of the total width. To further decrease the breathing amplitude we adjust the axial trap depth at the turning point of the breathing oscillation so that the shape of the instantaneous wavefunction matches that of the ground state wavefunction in the new trap.  This decreases the breathing amplitude by at least a factor of two.

After the trap turn-off and TOF expansion in each of our measurements, our data is collected via absorption imaging with a principal component analysis algorithm~\cite{PCA} to remove interference related features from the background.

To better measure the FWHM of the TOF distributions (see Figs.~\ref{fig:osc}, ~\ref{fig:osc2}, and ~\ref{fig:TheoryComparisons}) we apply a point smoothing function to reduce the noise without affecting the width of the distribution. 

\subsection{Dynamical Fermionization in a ``flat'' potential}
\begin{figure}[!hp]
	\includegraphics[width=0.95\textwidth]{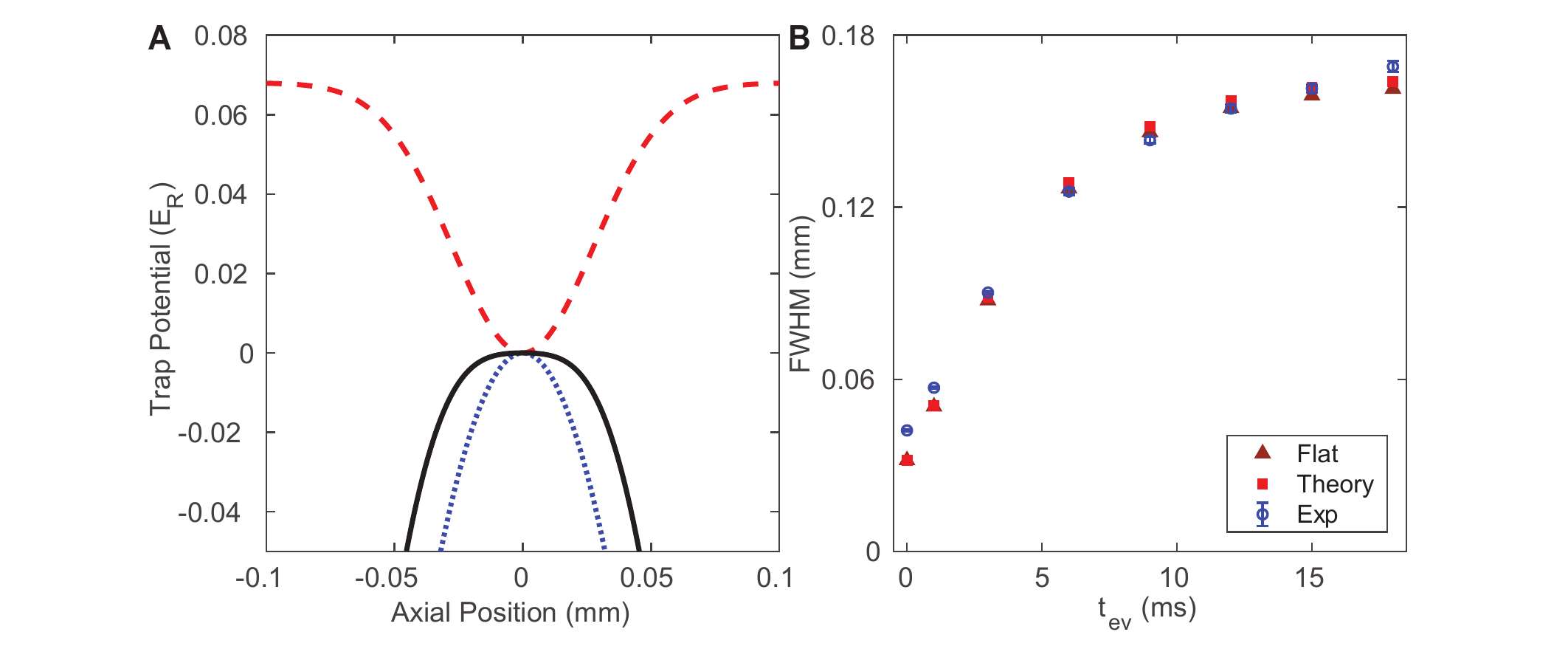}
	\caption{Cancellation of the anti-trap.(\textbf{A}) Anti-trapping potential due to the blue-detuned lattice light (shown in blue dots) and a shallow dipole trap (red dashed) are combined to form an effective ``flat'' potential (black solid) over a range of about $40$ $\mu$m during fermionization. (\textbf{B}) Full-width at half-maxima of the momentum distribution profiles from Fig.~\ref{fig:ferm} in the main text. By $t_{ev}=12$ ms the distributions have mostly fermionized but continue to broaden. Blue circles are the experimental widths, while red squares are the widths of profiles simulated in a potential that matches the experiment. The brown triangles are results from a simulation in a truly flat potential.  The slight discrepancy between squares and triangles is the residual effect of the anti-trap on the expansion, and is only visible at later times. The discrepancy between the experiment and the theory at the last point shows that the modeling of the potential gets worse too far from the center.}\label{fig:AntiTrap}
\end{figure}

A consequence of using blue-detuned light for transverse confinement is a weak anti-trap along the axial direction. If dynamical fermionization was done in the absence of any axial trapping light, the edges of the atomic cloud would accelerate away from the center due to the residual anti-trap, masking the contribution of the expansion from pure momentum evolution. To reduce the effect of the anti-trap we keep a shallow axial trap on for $t_{ev}$ after the quench in order to keep the potential ``flat’' over approximately $40\,\mu$m during expansion (see Fig.~\ref{fig:AntiTrap}A). Because the potential is not perfectly flat, both the experimental expansion and the theoretical expansion slightly deviate from theoretical expansion in a truly flat potential. The effect is small enough that it is only visible at late times (past 15 ms), as seen in Fig.~\ref{fig:AntiTrap}B. The disagreement between the experiment and the theory at the very last time in Fig.~\ref{fig:AntiTrap}B is due to a slight asymmetry in the Gaussian trap. All of our analysis is done before this asymmetry has affected our measurement.

\subsection{Lattice shutoff}

\begin{figure}[!hp]
	\includegraphics[width=1\textwidth]{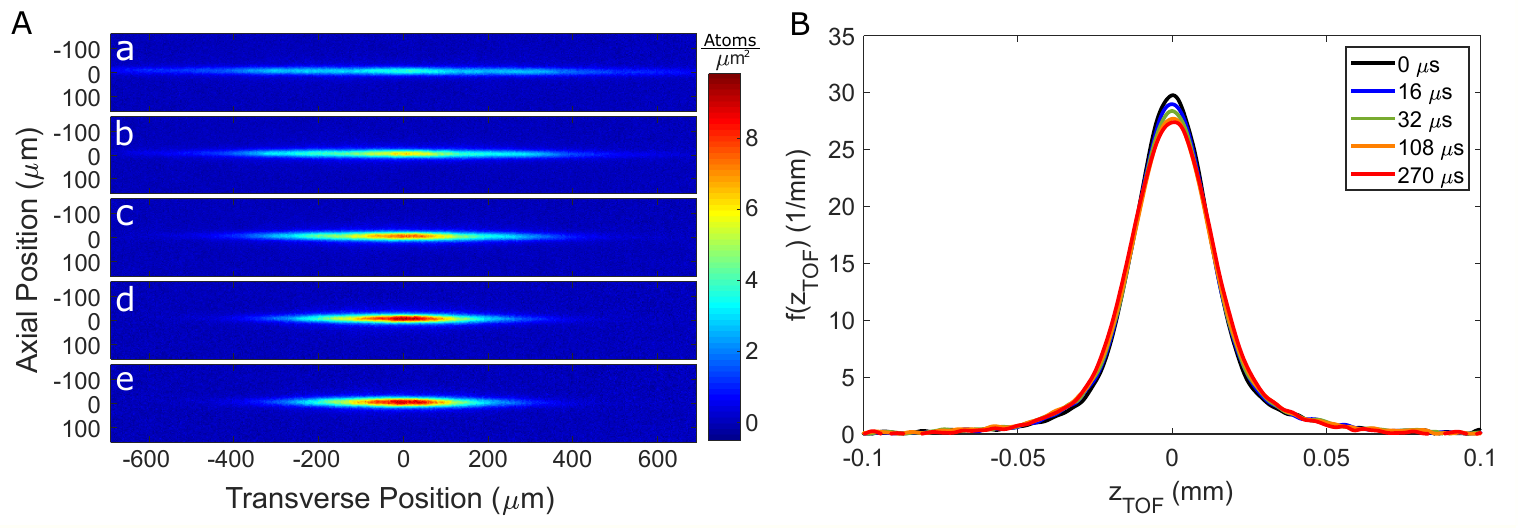}
	\caption{Lattice shutoff. (\textbf{A}) Absorption images at $t_{ev}=0$ taken with a $35\,$ms TOF, with different lattice shutoff speeds. (\textbf{a}), (\textbf{b}), (\textbf{c}), (\textbf{d}), and (\textbf{e}) are an average of 10 images for shutoff durations of $0,\,16,\,32,\,108$, and $270\,\mu$s respectively. As the shutoff becomes slower the transverse velocity of the cloud becomes smaller, allowing longer TOF for better momentum measurements. (\textbf{B}) Axial TOF distributions for the corresponding shutoffs. The shape changes in the form of slight broadening due to interactions during the shutoff time.}\label{fig:Shutoff}
\end{figure}

To turn off interactions and thus initiate a momentum measurement, we rapidly shut off the 2D lattice, which causes atoms to fly out transversely.  When atoms move too far past the imaging field of view, the magnetic field gradient they see is too small to exactly cancel gravity and the atoms fall slightly. Since atoms along the line of sight contribute to the integrated axial distributions, the TOF is limited by the extent of transverse expansion, which cannot be allowed too far past the imaging field of view. Fig.~\ref{fig:Shutoff}A shows the cloud after $35$ ms TOF with various lattice turn off times.  Longer turn off times lead to slower transverse expansion, but as can be seen in Fig.~\ref{fig:Shutoff}, longer shutoffs also lead to a slight change in the axial distribution due to the interaction energy. For dynamical fermionization we chose a shutoff that was $32\,\mu$s in duration. This speed is fast enough to prevent the distribution from changing dramatically and allows for a $70$ ms TOF without any apparent asymmetry from imperfect gravity cancellation. For the data with a quench from low to high $\omega_z$ the interactions are stronger, especially around the half-cycle mark, so we implement the sudden lattice shutoff and are only able to use a TOF of $40$ ms.  In the reverse quench, we use a $32\,\mu$s shutoff and $65$ ms TOF.

\subsection{Numerical Calculations} \label{method}

We model the experimental setup as an array of independent one-dimensional tubes described by the Lieb-Liniger Hamiltonian in the presence of a confining potential $V(z)$~\cite{olshanii_98}:
\begin{equation}\label{H_lieb_liniger}
{\cal H}_{\rm LL}=\sum_{j=1}^{N}\bigg[-\frac{\hbar^2}{2m}\frac{\partial^2}{\partial z^2_j}+V(z_j)\bigg]+g_{1D}\sum_{1\leq j < l \leq N}\delta(z_j-z_l) \,,
\end{equation}
where $m$ is the mass of the atoms, $N$ is the number of atoms, and $g_{1D}$ is the strength of the effective one-dimensional contact interaction. In the absence of $V(z)$, all observables depend only on the dimensionless quantity $\gamma={mg_{1D}}/{n_{1D}\hbar^2}$, where $n_{1D}$ is the particle density.

The number of particles in each tube depends on the $(x,y)$ position of the tube ~\cite{porto_04},
\begin{equation}\label{TF_distribution}
N(x,y)=N_{c}\left[1-\left(\frac{x}{R_{X}}\right)^2-\left(\frac{y}{R_{Y}}\right)^2\right]^{\frac{3}{2}}\,,
\end{equation}
where $N_c$ is the number of particles in the central tube, and $R_{X}$ ($R_{Y}$) is the Thomas-Fermi radius in the $x-$ ($y-$)direction. Given $R_{X}$, $R_{Y}$, and $N_\text{tot}$, which can be measured in the experiments, $N_c$ follows from $N_\text{tot}=\sum_{x,y}N(x,y)$.

All our numerical calculations are carried out in the Tonks-Girardeau limit ($\gamma\to\infty)$ in 1D lattices used to discritize space. The lattices are at very low fillings, so that the systems are effectively in the continuum (the average distance between particles is much larger than the lattice spacing). In this setup, the Bose-Fermi mapping allows one to compute correlation functions in inhomogeneous systems in equilibrium~\cite{rigol_muramatsu_04b, rigol_muramatsu_05b} and far from equilibrium~\cite{rigol_muramatsu_04d, rigol_muramatsu_05c} very efficiently. The corresponding lattice hard-core boson Hamiltonian has the form
\begin{equation}\label{H_HCB}
{\hat H}_{\rm HCB}=-J\sum_{j=1}^{L-1}\left(\hat b^{\dagger}_{j+1}\hat b^{}_{j}+\text{H.c.}\right)+\sum_{j=1}^{L}V(z_j)\hat b^{\dagger}_{j}\hat b^{}_{j}\,,
\end{equation}
where $\hat b^{\dagger}_j$ ($\hat b^{}_j$) denotes the creation (annihilation) of a hard-core boson at site $j$. Additional constraints $b^{\dagger 2}_j=b^{2}_j=0$ enforce the no-multiple-occupancy (hard-core) condition. In Eq.~\eqref{H_HCB}, $J$ is the hopping amplitude and $L$ is the total number of lattice sites. The system size in the continuum, $L_0$, is equal to $La$, where $a$ is the lattice spacing. The position of site $j$ in the lattice is taken to be $z_j=(j-L/2)a$.

One-body correlation functions of hard-core bosons are computed exactly via a mapping onto noninteracting spinless fermions using the Jordan-Wigner transformation, and then using properties of Slater determinants~\cite{rigol_muramatsu_04b, rigol_muramatsu_05b, rigol_muramatsu_04d, rigol_muramatsu_05c}. The parameters of the lattice Hamiltonian~(\ref{H_HCB}) and of the continuum Hamiltonian~(\ref{H_lieb_liniger}) satisfy the relation $J=\hbar^2/(2ma^2)$. As mentioned before, at low-fillings in the lattice (when $N/L\rightarrow0$), one obtains the same results as in the continuum \cite{xu_rigol_15}.

Our simulation of the experiments involves the following steps:

(i) The initial state in each tube is taken to be the ground state $|\psi_i\rangle$ with the appropriate particle number $N(x,y)$ (according to Eq.~\eqref{TF_distribution} rounded to the closest integer number) in the presence of a confining potential $V_\text{ini}(z)$. The latter is modeled as a sum of a Gaussian trap and a harmonic anti-trap
\begin{equation}\label{V_init}
V_\text{ini}(z)=U_\text{ini}\left[1-\exp\left(-\frac{2z^2}{W^2}\right)\right]-\frac{1}{2}m\omega_{at}^2 z^2\,,
\end{equation}
where $U_\text{ini}=(1/4)m\omega_\text{ini}^2W^2$ is the strength of the Gaussian trap, $W$ is the trap width, and $\omega_{at}$ is the anti-trapping frequency.

(ii) At $t=0$, we quench $V_\text{ini}\rightarrow V_\text{fin}$, where
\begin{equation}\label{V_final}
V_\text{fin}(z)=U_\text{fin}\left[1-\exp\left(-\frac{2z^2}{W^2}\right)\right]-\frac{1}{2}m\omega_{at}^2 z^2\,,
\end{equation}
The only parameter changed during the quench is the strength of the Gaussian trap, which becomes $U_\text{fin}=(1/4)m\omega_\text{fin}^2W^2$. We then compute the time evolution of the initial state under the final Hamiltonian,
$|\psi(t)\rangle=\exp(-iH_\text{fin}t/\hbar)|\psi_i\rangle$, and calculate the one-body density matrix $\rho_{lm}(t)=\langle\psi(t)|\hat b^{\dagger}_l \hat b^{}_m|\psi(t)\rangle$ and related observables such as the density (diagonal part) and the momentum (Fourier transform) distributions~\cite{rigol_muramatsu_04b, rigol_muramatsu_05b, rigol_muramatsu_04d, rigol_muramatsu_05c}.

(iii) After expansion for a time $t_{ev}$ in the presence of hard-core interactions in the one-dimension, we simulate the time-of-flight expansion for a time $\tau$ of the density distribution in the absence of interactions. For the dynamical fermionization $\tau=t_{det}-t_{ev}$, while for the oscillations $\tau=t_{TOF}$. The site occupations during time-of-flight expansion are given by the expression
\begin{equation}\label{free_prop}
n_j(\tau,t_{ev})=\sum_{m,n}G^*_{j,l}(\tau)G_{j,m}(\tau)\rho_{lm}(t_{ev})\,,
\end{equation}
where $G_{m,n}(\tau)=\sum_k \exp\{-i\tau/\hbar[\epsilon_k-\hbar k(z_m-z_n)/\tau]\}$ is the free one-particle propagator with a dispersion relation $\epsilon_k=-2J\cos(ka)$~\cite{rigol_muramatsu_05a} .

(iv) We then sum the contributions from all tubes, and properly normalize density and momentum distributions so that the area under those distributions is one.

(v) Finally, for the normalized time-of-flight density distributions $\bar{n}(z)$, we use a Gaussian function $R(z)=\exp[-z^2/(2\sigma^2)]/(\sqrt{2\pi}\sigma)$ to convolve the exact numerical results with the resolution of the imaging system. The result of that convolution is what we compare to the experimental results.

We use a lattice spacing $a=4\times10^{-8}$ m for the expansion calculations that reproduce the dynamical fermionization of the momentum distribution, and $a=2\times10^{-8}$ m for the calculations of the Bose-Fermi oscillations. The total number of lattice sites for the time evolution in the presence of hard-core interactions is $L=3000$, while the time-of-flight calculations are carried out in lattices with up to $50000$ sites.

\subsection{Quench from low to high $\omega_z$}

\begin{figure}[!hp]
	\includegraphics[width=1\textwidth]{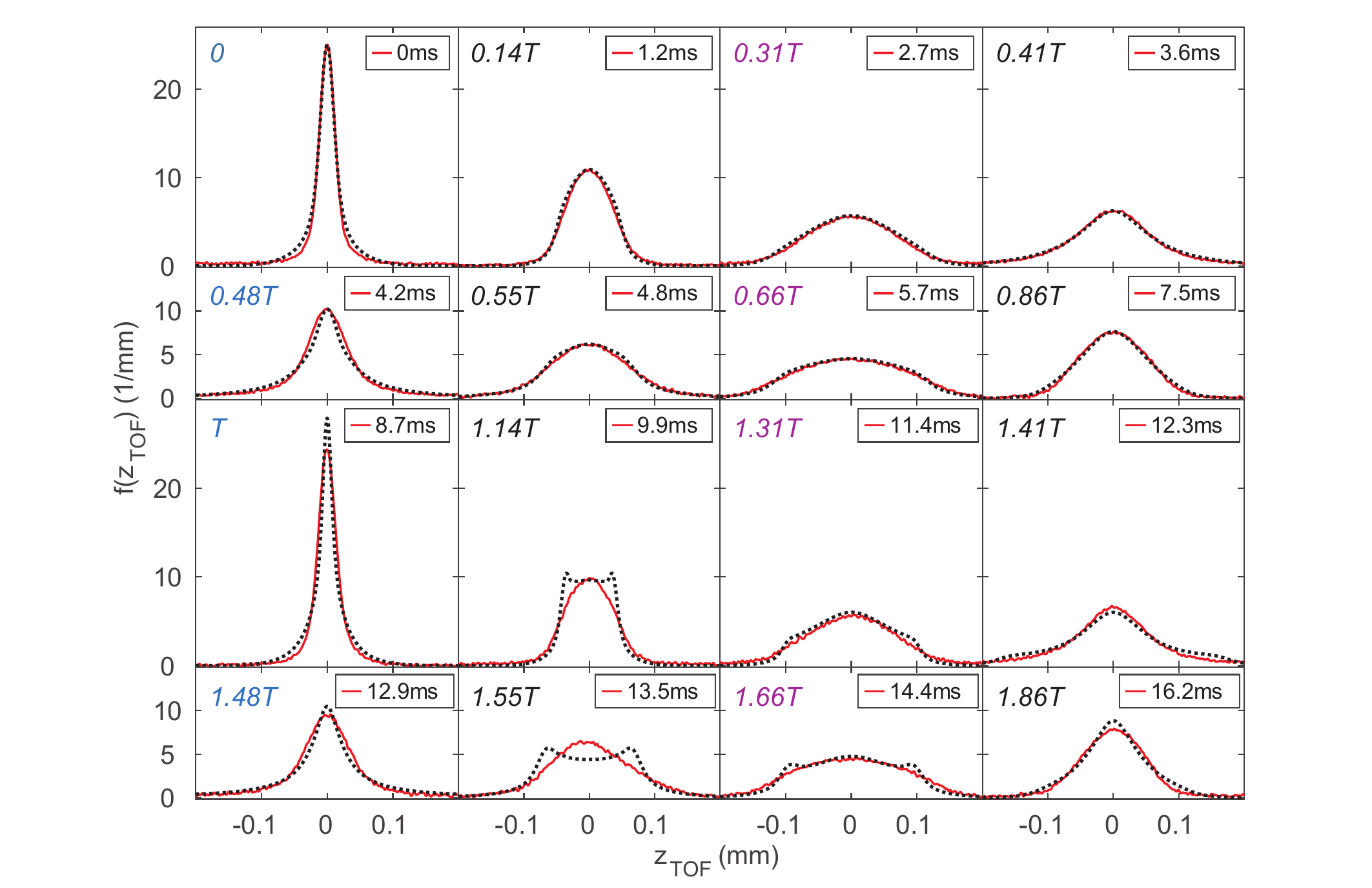}
	\caption{Bosonic-Fermionic oscillation (quench low to high $\omega_z$). The solid lines are experimental TOF distributions, and the dotted lines are rescaled and shifted simulation results. The times associated with each profile are the $t_{ev}$ for the experimental distributions.  The labeled fractions of the oscillation period $T$ refer to both experiment and theory.  The times marked in blue are near the FWHM minima (bosonic distributions) and the times marked in red are near the FWHM maxima (fermionic distributions). The intermediate times also appear fermionic in shape.  In the first period the shapes agree very well, and in the second period the experimental shapes basically repeat.  The theory in the second period has sharper features that only appear for an anharmonic trap. That they are absent from the experiment is a consequence of reduced $\gamma$. }\label{fig:EBP}
\end{figure}

The ratio of the theoretically simulated breathing oscillation frequency to the dipole frequency for a T-G gas in a harmonic trap is 2, but we observe $\sim1.96$ because of the anharmonic (Gaussian) trap. More significantly, the quench from low to high $\omega_z$ after a 10-fold increase in trap depth causes the density to increase substantially. Therefore the initial average $\gamma$ of 8.5 decreases over the course of the oscillation, reaching as low as $\sim$2. The equilibrium $\gamma$ in the new trap would be close to 4, but the atoms compress past the equilibrium axial size. We observe that the experimental breathing period is 9$\pm$0.6$\%$ smaller than the T-G theory, which we ascribe to the departure from the T-G regime  (Figs.~\ref{fig:osc} and~\ref{fig:EBP}).  When $\gamma$ decreases from $\infty$ to 0, the breathing period is expected to decrease by 13.4$\%$ (a reduction factor of $\sqrt{3}/2$)~\cite{menotti_stringari_02, moritz_stoferle_03}. We have not attempted a detailed comparison of the experiment to a finite $\gamma$ theory in an anharmonic trap, but the observation is roughly in line with expectations, since the theoretical halfway point in the breathing period (a 6.7$\%$ reduction from the T-G gas) occurs at $\gamma\sim3$ (see Fig. 3 in Ref.~\cite{menotti_stringari_02}).

The reduced $\gamma$ after this quench also affects the momentum distributions. Fig. \ref{fig:EBP} shows comparisons between the experiment and theory at a range of times (see also Fig. \ref{fig:osc}). The first two rows are selected from the first period of oscillation. We compare profiles that have the same phase in their oscillation, not the same absolute time. In order to compare the shapes of the broader theoretical distributions to the narrower experimental ones, the theory has been rescaled to match the height in the data while keeping the area constant. Despite the finite $\gamma$, the shapes in the first period of oscillation agree very well. The narrowest profiles are bosonic in shape, while the rest of the points are predominantly fermionic in shape. That is, while the details of the breathing (period and widths) differ from the T-G gas, the qualitative bosonic-fermionic oscillations are quite insensitive to the reduced $\gamma$.

\begin{figure}[!hp]
	\includegraphics[trim={0 25 0 35},width=.75\textwidth]{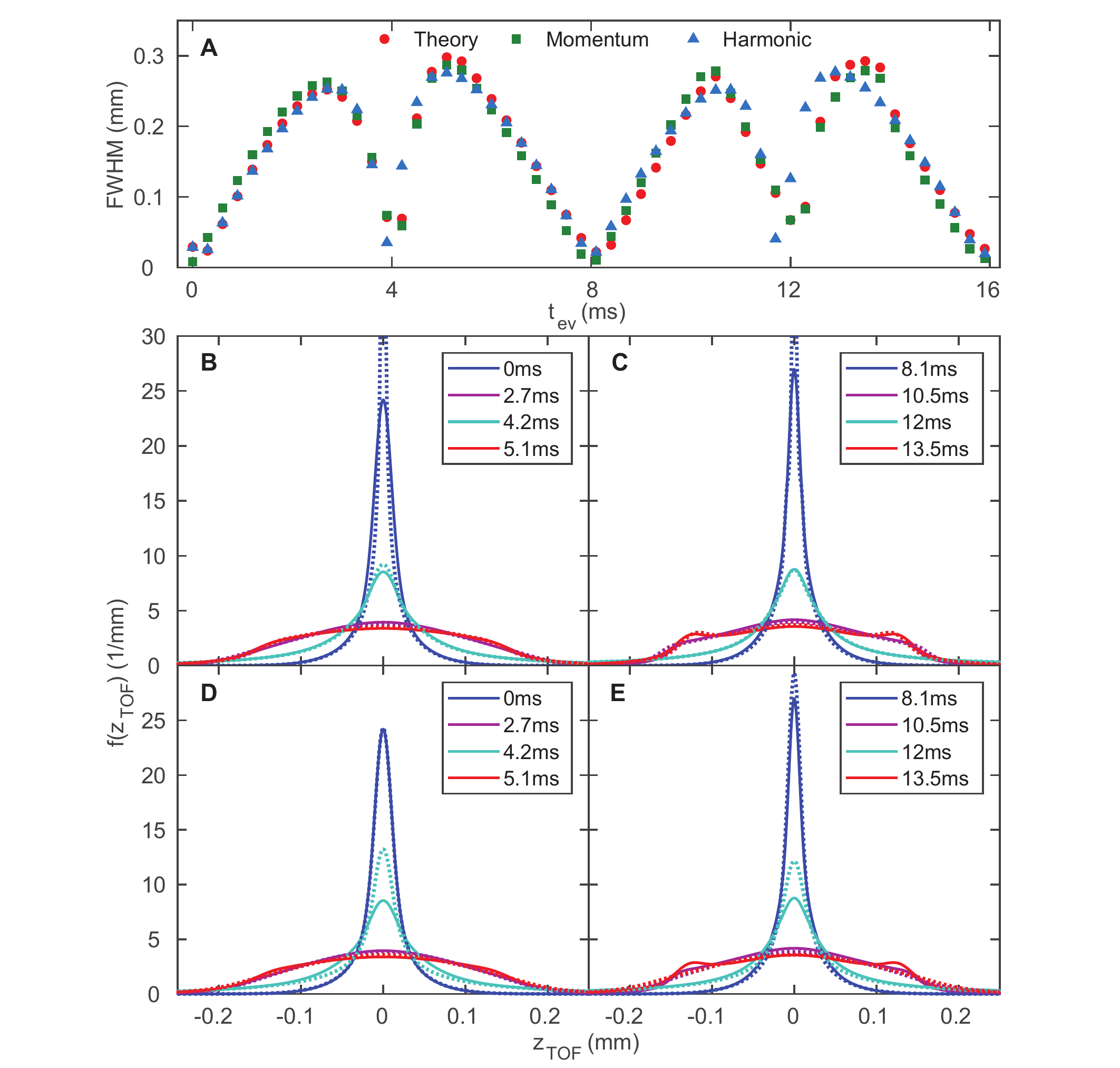}
	\caption{Comparisons of different versions of the theoretical TOF calculations, for the quench from low to high $\omega_z$. (\textbf{A}) FWHM vs. time.  The red circles represent the TOF theory for a Gaussian trap. The green squares refer to the corresponding momentum distributions, rescaled by the TOF so they can be compared to the red circles, to which they closely correspond. For a few points in the second period, the center of the distribution is not the maximum (see Fig~\ref{fig:EBP}); in those cases we still define the half maximum relative to the center point. The blue squares represent the TOF theory in a harmonic trap. The second cycle looks the same as the first for this curve, unlike the curves associated with a Gaussian trap. (\textbf{B}) Distributions from the first period. The Gaussian TOF distributions (solid lines) and the scaled momentum distributions (dotted lines) nearly overlap except at the earliest times, when the initial size affects the TOF distributions. (\textbf{C}) Distributions from the second period. The Gaussian TOF curves (solid lines) and the scaled momentum distributions (dotted lines) are still similar except when the distributions are narrowest. (\textbf{D}) Distributions from the first period. The Gaussian TOF distributions (solid lines, same data as in \textbf{B}) and the simulation in a harmonic trap (dotted lines) are qualitatively similar, although the widths differ at the half cycle point. (\textbf{E}) Distributions from the second period. The simulation in a harmonic trap (dotted lines) do not have any of the sharp features that are in the Gaussian TOF distributions (solid lines, same data as in \textbf{C}). }\label{fig:TheoryComparisons}
\end{figure}

The last two rows in Fig. \ref{fig:EBP} are from the second period. Because the theoretical distributions in the second period have more complex shapes, we use the rescaling factors from the corresponding points in the first period. In the second period the experimental data basically repeats itself, while the theory starts to show sharper features.  These disappear when the theory is simulated in a harmonic instead of a Gaussian trap (Fig.~\ref{fig:TheoryComparisons}). Apparently, the reduced $\gamma$ softens the sharp features; similar behavior has been seen in other hard-core boson theory/experiment comparisons \cite{Vidmar_15}.

As noted earlier, the increased density in the quenched trap forces us to use a faster shutoff and shorter TOF for this measurement than for the dynamical fermionizations measurement.  Fig.~\ref{fig:TheoryComparisons}B and Fig.~\ref{fig:TheoryComparisons}C show that despite this short time of flight, the TOF distributions we measure are close approximations of the actual momentum distributions in the trap. The exception is the earliest time, where the initial size of the cloud noticeably affects the TOF.

 The wide momentum distributions that occur during the oscillations after this quench cause the signal to spread out over a larger part of the CCD than in our other measurements. This causes our principal component analysis background subtraction~\cite{PCA} to become less reliable. Compromised background subtraction is visible in the CCD regions where there are no atoms; if the background deviates from zero by more than 10$\%$ of the peak value of the widest distribution ($t_{ev}$=5.7 ms), we reject that image.  For this particular quench only, this leads us to reject between 1 and 10 of the 15 shots taken at each $t_{ev}$ .

\subsection{Quench from high to low $\omega_z$}

\begin{figure}[!hp]
	\includegraphics[width=1\textwidth]{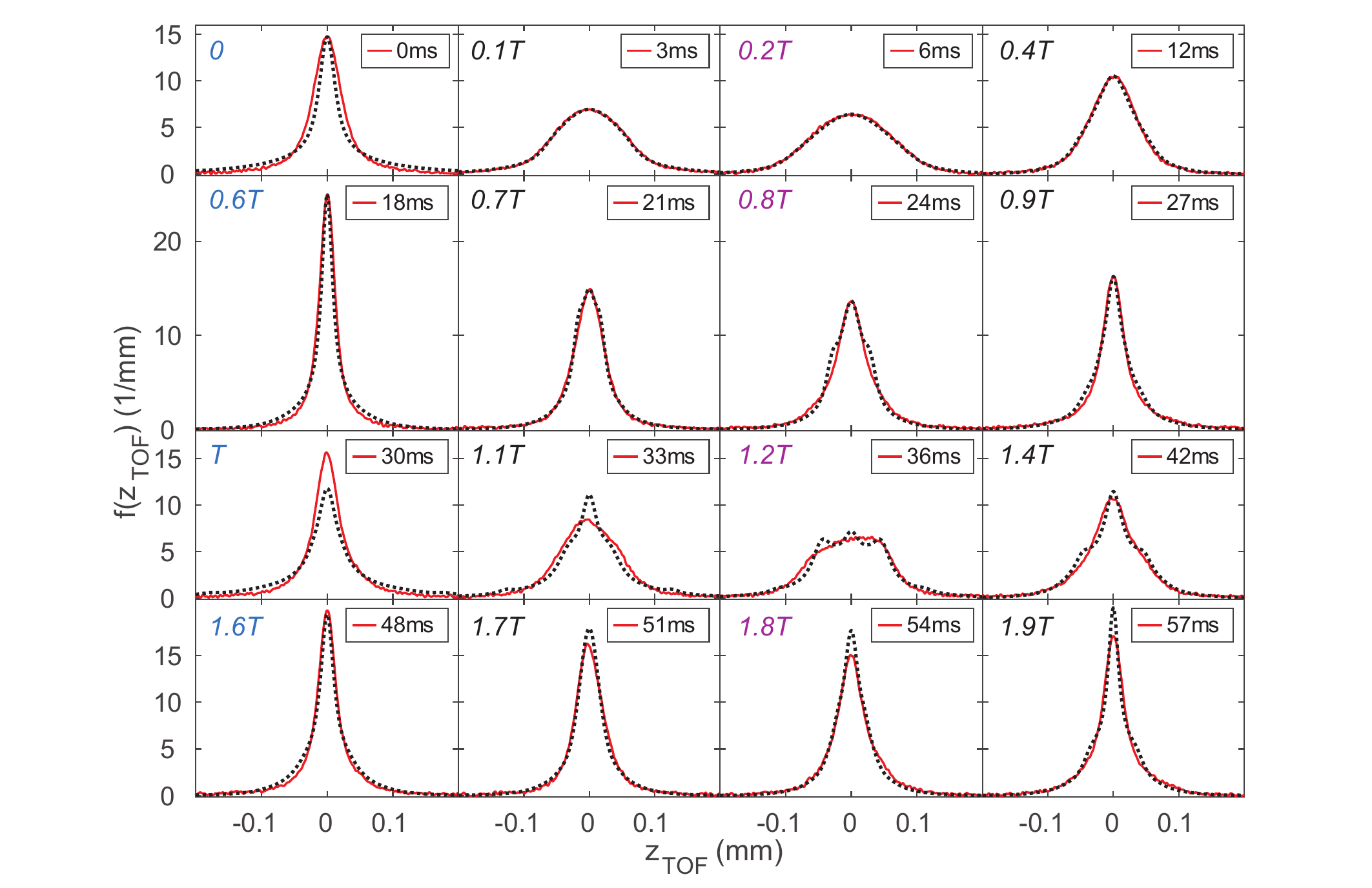}
	\caption{Bosonic-Fermionic oscillation (quench high to low $\omega_z$). The solid lines are experimental TOF distributions, and the dotted lines are rescaled simulation results. The absolute time and the fractions of the oscillation period $T$ refer to both experiment and theory. The times marked in blue are near the FWHM minima (bosonic distributions) and the times marked in red are near the FWHM maxima (fermionic distributions). The theory once again shows peaked features due the Gaussian trap, and the experiment shows evidence of such features, particularly at $1.2\,T$.}\label{fig:EBP2}
\end{figure}

For the quench from high to low frequency (3$\times$ decrease in trap depth), we start with a higher density of atoms confined in fewer tubes than in the previously discussed measurements. After the quench, an initial average $\gamma$ of 4.4  reaches $\sim$6.7 during the oscillation. There is essentially no frequency difference between the theoretical and experimental oscillations (see Figs.~\ref{fig:osc2} and~\ref{fig:EBP2}). We expect that the finite $\gamma$ should decrease the frequency as in the other quench, but not by quite as much (we estimate 6$\%$)~\cite{menotti_stringari_02}. That effect, however, seems to be counterbalanced by another. The ratio T-G gas theory breathing frequency to the dipole trap frequency is 1.63, but in a harmonic trap this ratio would be 2. Since the fermionic momentum distributions are broader in the theory than in the experiment, the atoms in the experiment do not spread as far in the Gaussian trap, so they will have a relatively higher breathing frequency due to this effect. Here again, theoretical calculations at finite $\gamma$ in anharmonic traps will be needed to quantitatively support this explanation.

As just noted, we see narrower FWHM for fermionic experimental distributions than for the corresponding T-G gas theoretical distributions. In order to compare the shapes, we therefore rescale the theoretical curves in Fig.~\ref{fig:EBP2} as we did in Fig.~\ref{fig:EBP}. Unlike in the other quench, for this quench to higher $\gamma$, the experimental curves slightly change in the second period compared to the first. For instance, the 1.2$T$ peak is flattened, and the 1.4$T$ curve has small shoulders. There are somewhat sharper features in the associated theory curves, which otherwise remain similar throughout. The experimental behavior, especially in the second period, tends to confirm our association with finite $\gamma$ of the loss of the sharp features due to trap anharmonicity. (Note that the height difference between the experiment and theory at the beginning of the second period occurs because we rescale with the same factor as in the first cycle. Both curves look similarly bosonic.)

 We use a three times smaller quench in trap depth ($\sqrt{3}$ in $\omega_z$) than when we quench to a deeper trap in order to limit the spatial extent of the trapped atoms. When they are too spread out we start to see the effect of dipole beam imperfections. The slight asymmetry in the 1.2$T$ experimental curve in Fig.~\ref{fig:EBP2} hints at this limitation. It is remarkable that even with such a small quench, we still see reasonably clear bosonic-fermionic oscillations.

\newpage
\section*{Figures S1-S4}

\newpage
\newpage
\newpage
\newpage
\newpage
\clearpage

\end{document}